\documentclass[preprint,prd,tightenlines,floatfix,
nofootinbib,eqsecnum,superscriptaddress]{revtex4}

\usepackage[T1]{fontenc}		

\usepackage{soul}

\usepackage{amsmath,amsfonts,amssymb,amstext,mathrsfs}
\usepackage{amsthm}
\usepackage{mathpazo}

\usepackage[dvips]{graphicx}
\usepackage{epsf,float}
\usepackage{revsymb}

\usepackage{dcolumn}
\usepackage{color,xcolor}
\usepackage{graphicx}
\usepackage{tabularx}
\usepackage{pstricks}
\usepackage[section]{placeins}
\usepackage{booktabs}
\usepackage{array}

\usepackage{hyperref}

\usepackage{slashed}

\begin{document}

\author{Rafa{\l} Maciu{\l}a}
\email{Rafal.Maciula@ifj.edu.pl}
\affiliation{Institute of Nuclear Physics Polish Academy of Sciences, 
	Radzikowskiego 152, PL-31342 Krak\'ow, Poland}

\author{Wolfgang Sch\"afer}
\email{Wolfgang.Schafer@ifj.edu.pl}
\affiliation{Institute of Nuclear Physics Polish Academy of Sciences, 
	Radzikowskiego 152, PL-31342 Krak\'ow, Poland}

\author{Antoni Szczurek
	\footnote{Also at \textit{College of Natural Sciences, 
			Institute of Physics, University of Rzesz\'ow, 
			Pigonia 1, PL-35310 Rzesz\'ow, Poland}.}}
\email{Antoni.Szczurek@ifj.edu.pl}
\affiliation{Institute of Nuclear Physics Polish Academy of Sciences, 
	Radzikowskiego 152, PL-31342 Krak\'ow, Poland}

\title{On the mechanism of $T_{4c}(6900)$ tetraquark production}

\begin{abstract}
We discuss the production mechanism of a new state, 
a putative fully charm tetraquark, observed recently by the LHCb at 
M = 6.9 GeV in the $J/\psi J/\psi$ channel.
Both single parton scattering (SPS) and double parton scattering (DPS)
mechanisms are considered. We calculate the distribution in the invariant 
mass of the four-quark system $M_{4c}$ for SPS and DPS production of 
$c c \bar c \bar c$ in the $k_t$-factorization approach with modern
unintegrated gluon distribution functions (UGDFs). 
The so-calculated contribution of DPS is almost two orders of 
magnitude larger than the SPS one, but the tetraquark formation mechanism
is unknown at present.
Imposing a mass window around the resonance position we calculate the
corresponding distribution in $p_{t,4c}$ -- the potential tetraquark
transverse momentum. The cross section for the $J/\psi J/\psi$ continuum
is calculated in addition, again including SPS (box diagrams) and 
DPS contributions which are of similar size.
The formation probability is estimated trying to reproduce
the LHCb signal-to-background ratio. The calculation of the SPS
$g g \to T_{4c}(6900)$ fusion mechanism is performed in the
$k_T$-factorization approach assuming different spin scenarios ($0^+$
and $0^-$). The $0^+$ assignment is preferred over the $0^-$ one
by the comparison of the transverse momentum distribution of signal
and background with the LHCb preliminary data assuming the SPS mechanism
dominance. There is no reliable approach for the DPS formation mechanism
of tetraquarks at present as this is a complicated multi-body problem.
\end{abstract}

\maketitle

\section{Introduction} 

The potential existence of tetraquarks was discussed a decade after 
the quark model was formulated \cite{Jaffe}. 
Although the potential 
territory of tetraquarks is large there is an ongoing debate regarding their 
identification. The conjectured light tetraquarks, like e.g. $f_0(980)$, 
are ambiguous, because often a competitive 
interpretation as a hadronic molecule ($K \bar K$ for $f_0(980)$) or 
as a more generic coupled-channel/threshold phenomenon is possible. 

Much attention has recently been paid to possibly exotic hadrons
containing heavy quarks, where a plethora of newly discovered states 
await their definitive theoretical understanding/interpretation \cite{Esposito:2016noz,Karliner:2017qhf,Olsen:2017bmm,julich}.

The recent observation by the LHCb collaboration \cite{LHCb_T4c} of a sharp peak in the di-$J/\psi$
channel at $M$ = 6.9 GeV seems to strongly suggest the presence of 
a fully charm tetraquark, consisting of $c c \bar c \bar c$.

During the last years a number of theoretical models for the spectroscopy of 
tetraquarks were developed and are waiting for experimental verification.
The most popular approach treats the fully heavy ($c c \bar c \bar c$,
$b b \bar b \bar b$ or $c \bar c b \bar b$) tetraquarks as a bound system
of a color antitriplet diquark and color triplet antidiquark.
In early works the diquarks were treated as structureless
objects, and the interaction between diquark and antidiquark is then 
constructed in analogy to that between heavy quarks developed earlier 
for quarkonia \cite{KNR2017,lebed2017,DN2019,BFRS2019,GL2020}. The second color configuration of a
sextuplet diquark and antisextuplet antidiquark is most often neglected. 
Reservations regarding the diquark approach from the point of view of 
more rigorous approaches to the few-body problem have been raised 
in \cite{Richard:2018yrm}.

Assuming the enhancement observed by LHCb is indeed caused by a new
state, models suggest that it is rather an excited state.
There is no clue at present on its spin and parity. Clearly, higher
statistics studies which may give model-independent answer
\cite{MAN2020} are required in future. Different models predict slightly
different quantum numbers. For example the nonrelativistic potential 
quark model (NRPQM) \cite{LLZZ2020} predicts the state to be 
$J^{PC}$ = $0^{-+}$ or $1^{-+}$. A similar result is obtained in 
the framework of a QCD sum rule approach \cite{CCLZ2020}. 
A different pattern is obtained in the relativized quark
model with quark-quark, antiquark-antiquark and quark-antiquark
interactions \cite{LCD2020}. In this approach the 6.9 GeV state can be
a radial excitation with $J^{PC}$ = $0^{++},2^{++}$.
A caveat regarding the interpretation of the LHCb result is in order: 
in \cite{WCLM2020} the authors describe the di-$J/\psi$
invariant mass distribution by a non-resonant rescattering of
quarkonium pairs. The peaks are then associated with threshold
enhancements for different channels.

With this reservation in mind, we will in the following assume that indeed
an (excited) tetraquark state has been observed.
Different $J^{PC}$ combinations are possible in general
\cite{DN2019,BFRS2019,CCLZ2020,LCD2020}, the details depend
on the method used for the the four-body systems.

The decays of the fully heavy tetraquarks was discussed e.g. 
in \cite{CCLZ2020,BGMS2006}.

On the other hand the production mechanism of the $T_{4c}$ tetraquarks 
is terra incognita.
There are only a few papers \cite{BLLN2011,CCGN2016} on the
production of the ground-state fully charm tetraquark. 
The production cross section for $T_{4c}$ in \cite{CCGN2016}
was estimated to be one order of magnitude smaller than that for $X(3872)$
(assumed to be a tetraquark) production which was measured by 
the CMS collaboration, the production mechanism of $X(3872)$ however 
itself is under debate. 

What is the mechanism of the fusion of four charm
quarks/antiquarks is not clear at the moment. 

Some time ago a large cross section for 
$c \bar c c \bar c$ production at the LHC due to
double-parton scattering (DPS) mechanism was predicted in
\cite{LMS2012}.
The SPS mechanism was also considered but its contribution to the $c \bar c c \bar c$ production is much smaller 
\cite{SS2012,MS2013,HMS2014,HMS2015}.
The prediction of Ref. \cite{LMS2012} was verified by the LHCb collaboration by observing many
same-flavor $D$ mesons \cite{LHCb_DD}. The calculation in the 
$k_T$-factorization approach explained many correlation observables
for double $D$ meson production \cite{MS2013,HMS2014,HMS2015}.

The production of two $J/\psi$ states was also studied at the
LHC \cite{LHCb_jpsijpsi,ATLAS_jpsijpsi,CMS_jpsijpsi}. The situation at 
the LHCb kinematics seems better understood 
\cite{KKS2011,Baranov:2011zz,BSZSS2013,BK2017} than for the ATLAS or 
CMS one where the number of crucial mechanisms seems much bigger
(see e.g. \cite{He:2015qya,Lansberg:2014swa,Lansberg:2019fgm,He:2019qqr,SCS2017}).

Recently, within $k_T$-factorization, also the gluon-gluon fusion 
mechansism of production of pseudoscalar \cite{babiarz_pseudoscalar} 
and scalar \cite{babiarz_scalar} charmonia was studied. 
Here we shall consider the mechanism of gluon-gluon fusion
for the fully charm tetraquark production. In this letter we shall
discuss only $0^+$ and $0^-$ scenarios where the formalism was tested.

\section{Cross section for signal and background}

In this section we discuss several issues related to the production
of the $T_{4c}(6900)$ tetraquark.

After many years of investigation there is no agreement on production 
mechanism even for quarkonia, pure $Q \bar Q$ states.
For $C$ = +1 quarkonia rather color singlet mechanism dominates 
\cite{babiarz_pseudoscalar}. How big is color octet contribution
is not quite clear at present.

The reaction mechanism for $C$ = + 1 tetraquark production (the LHCb case) 
can be categorized as:\\
(a) $c \bar c c \bar c$ are produced in color singlet state, \\
(b) $c \bar c c \bar c$ are produced in color octet state
and extra emission(s) of soft gluon(s) is(are) necessary to bring
the $c \bar c c \bar c$ system to color singlet state relevant for
the tetraquark hadron.

\subsection{$p p \to c \bar c c \bar c$ cross section}

\begin{figure}
\includegraphics[width=5cm]{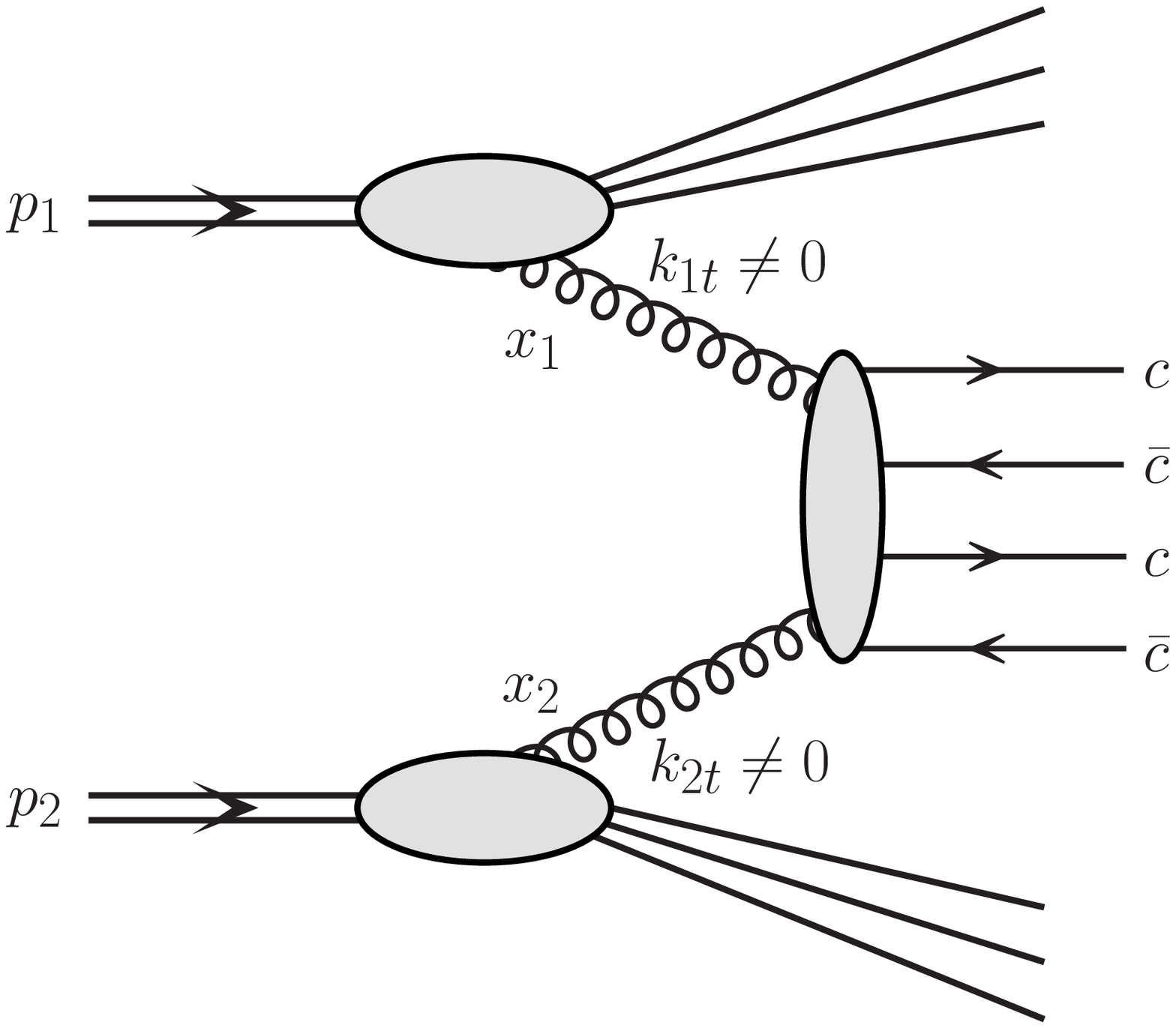}
\includegraphics[width=5cm]{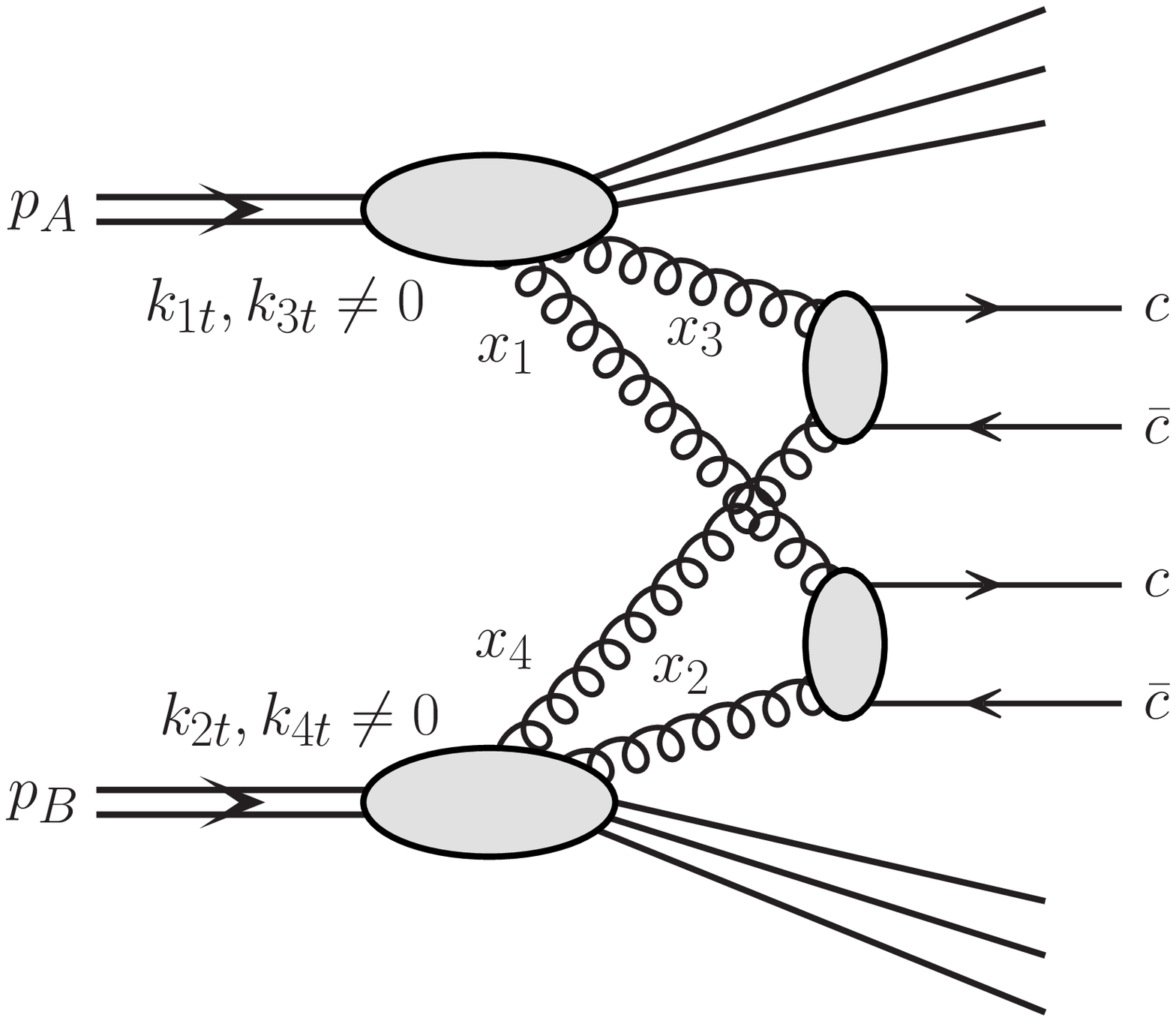}
\caption{Two dominant reaction mechanisms of production of 
$c \bar c c \bar c$ nonresonant continuum. The left diagram represent
the SPS mechanism (box type) and the left diagram the DPS mechanism.
}
\label{fig:diagrams_ccbarccbar}
\end{figure}

In this subsection we wish to calculate the cross section for
four charm quark/antiquark production. In particular, we wish to calculate
distribution in invariant mass of the four charm quarks/antiquarks
in the region of low invariant masses. In particular, such a cross
section in the mass window arround the mass of the tetraquark can 
be compared to the cross section for the tetraquark which at present 
can be only estimated with poor precision.
In Fig.\ref{fig:diagrams_ccbarccbar} we show the dominant reaction
mechanisms: SPS type (left diagram) and DPS type (right diagram).

In the present study both the SPS and the DPS contributions are
calculated in the framework of $k_{T}$-factorization 
\cite{Catani:1990xk,Catani:1990eg,Collins:1991ty,Gribov:1984tu}.
According to this approach the SPS cross section for 
$pp \to c\bar c c\bar c \, X$ reaction can be written as
\begin{equation}
d \sigma_{p p \to c\bar c c\bar c \; X} =
\int d x_1 \frac{d^2 k_{1t}}{\pi} d x_2 \frac{d^2 k_{2t}}{\pi}
{\cal F}_{g}(x_1,k_{1t}^2,\mu^2) {\cal F}_{g}(x_2,k_{2t}^2,\mu^2)
d {\hat \sigma}_{g^*g^* \to c\bar c c\bar c}
\; .
\label{cs_formula}
\end{equation}
In the formula above ${\cal F}_{g}(x,k_t^2,\mu^2)$ is the unintegrated or transverse momentum dependent
gluon distribution function (gluon uPDF). The uPDF depends on longitudinal momentum fraction $x$, transverse momentum squared $k_t^2$ of the partons entering the hard process,
and in general also on a (factorization) scale of the hard process $\mu^2$.
The elementary cross section in Eq.~(\ref{cs_formula}) can be written as:
\begin{equation}
d {\hat \sigma}_{g^*g^* \to c \bar c c \bar c} = { 1 \over (2!)^2}
\prod_{l=1}^{4}
\frac{d^3 \vec p_l}{(2 \pi)^3 2 E_l} 
(2 \pi)^4 \delta^{4}(\sum_{l=1}^{4} p_l - k_1 - k_2) \frac{1}{\mathrm{flux}} \overline{|{\cal M}_{g^* g^* \to c \bar c c \bar c}(k_{1},k_{2},\{p_l\})|^2}
\; ,
\label{elementary_cs}
\end{equation}
where $E_{l}$ and $p_{l}$ are energies and momenta of final state charm quarks. 
The matrix element takes into account that both gluons entering the hard
process are off-shell with virtualities $k_1^2 = -k_{1t}^2$ and $k_2^2 = -k_{2t}^2$.
In numerical calculations we limit ourselves to the dominant gluon-gluon fusion channel of the $2 \to 4$ type parton-level mechanism.
We checked numerically that the channel induced by the $q\bar q$-annihilation can be safely neglected in the kinematical region under consideration here.

A formal theory of multiple-parton scattering (see \textit{e.g.}
Refs.~\cite{Diehl:2011tt,Diehl:2011yj}) is rather well established but still not fully applicable for phenomenological studies. In general, the DPS cross sections can be expressed in terms of the double parton distribution functions (dPDFs).
However, the currently available models of the dPDFs are still rather 
at a preliminary stage. So far they are formulated only for gluon or for valence quarks and only in a leading-order framework
which is for sure not sufficient for many processes, especially when heavy quark production is considered.    

In general, the DPS cross sections can be expressed in terms of the double parton distribution functions (dPDFs) (see \textit{e.g.}
Refs.~\cite{Diehl:2011tt,Diehl:2011yj}). However, the currently available models of the dPDFs are still rather 
at a preliminary stage. Therefore, in phenomenological studies one usually follows the assumption of the factorization of the DPS cross section.
Within the factorized ansatz, the dPDFs are taken in the following form:  
\begin{equation}
D_{1, 2}(x_1,x_2,\mu) = f_1(x_1,\mu)\, f_2(x_2,\mu) \, \theta(1-x_1-x_2) \, ,
\end{equation}
where $D_{1, 2}(x_1,x_2,\mu)$ is the dPDF and
$f_i(x_i,\mu)$ are the standard single PDFs for the two generic partons in the same proton. The factor $\theta(1-x_1-x_2)$ ensures that
the sum of the two parton momenta does not exceed 1. 

According to the above, the differential cross section for $pp \to c \bar c c \bar c \; X$ reaction within the DPS mechanism can be then expressed as follows: 
\begin{equation}
\frac{d\sigma^{DPS}(pp \to c \bar c c \bar c \;X)}{d\xi_{1}d\xi_{2}} =  \frac{m}{\sigma_{\mathrm{eff}}} \cdot \frac{d\sigma^{SPS}(pp \to c \bar c \;X)}{d\xi_{1}} \! \cdot \! \frac{d \sigma^{SPS}(pp \to c \bar c \;X)}{d\xi_{2}},
\label{basic_formula1}
\end{equation}
where $\xi_{1}$ and $\xi_{2}$ stand for generic phase space kinematical variables for the first and second scattering, respectively.
The combinatorial factor $m$ is equal $0.5$ for the $c \bar c c\bar c$ case. Here, the $d\sigma^{SPS}(pp \to c \bar c \;X)$ ingredient cross sections are also calculated with the off-shell initial state partons.
%
%

The effective cross section $\sigma_{\mathrm{eff}}$ provides normalization of 
the DPS cross section and can be roughly interpreted 
as a measure of the transverse correlation of the two partons inside 
the hadrons. The longitudinal parton-parton correlations are far less
important when the energy of the collision is increased, due to the
increase in the parton multiplicity. For small-$x$ partons and for low 
and intermediate scales the possible longitudinal correlations can be safely
neglected (see \textit{e.g.} Ref.~\cite{Gaunt:2009re}). 
In this paper we use world-average value of $\sigma_{\mathrm{eff}} = 15$ mb provided by 
several experiments at Tevatron
\cite{Abe:1997bp,Abe:1997xk,Abazov:2009gc} and LHC
\cite{Aaij:2012dz,Aad:2013bjm,Chatrchyan:2013xxa,Aad:2014rua,Aaboud:2016dea}.
Future experiments may verify this value and establish a systematics.

The numerical calculations for both the SPS and the DPS mechanisms 
are performed with the help of KaTie \cite{vanHameren:2016kkz}, which is 
a complete Monte Carlo parton-level event generator for hadron
scattering processes. 
It can can be applied to any arbitrary processes within the Standard
Model, for several final-state particles, and for any initial partonic
state with on-shell or off-shell partons.
We use $\mu^2 \! = \! \sum_{i=1}^{4} m_{it}^{2}/4$ as the renormalization/factorization scale, where $m_{it}$'s are the transverse masses of the outgoing charm quarks. We take running $\alpha_{s}$ at next-to-leading order (NLO) and
charm quark mass $m_c$ = 1.5 GeV. Uncertainties related to the choice of the parameters were
discussed very recently in Ref.~\cite{Maciula:2017egq} and will be not considered here. We use the Kimber-Martin-Ryskin (KMR) \cite{Kimber:2001sc,Watt:2003vf} unintegrated distributions for gluon calculated from the MMHT2014nlo PDFs \cite{Harland-Lang:2014zoa}. The above choices are kept the same also in the case of double-parton scattering calculation except of the scales. 

Having calculated differential cross section for $c\bar c c\bar c$-system
production one can obtain the cross section for $T_{4c}(6900)$ tetraquark
within the framework of color evaporation model (CEM)
\cite{MV2016,CV2017}. The $c\bar c c\bar c \to T_{4c}(6900)$ transition can be
written as follows:
\begin{eqnarray}
\frac{d\sigma_{T_{4c}} }{d^3\vec P_{T_{4c}}} = F_{T_{4c}}
 \int_{M_{T_{4c}}-\Delta M}^{M_{T_{4c}}+\Delta M} d^3 \vec{P}_{4c} \; d
 M_{4c} \frac{d\sigma_{c\bar c c\bar c}}{ d M_{4c} d^3 \vec P_{4c}}
 \delta^3(\vec{P}_{T_{4c}}-\frac{M_{T_{4c}}}
{M_{4 c}} \vec{P}_{4c}),
\end{eqnarray}
%
where  $F_{T_{4c}}$ is the probability of the $c \bar c c \bar c \to T_{4c}$ transition which is unknown and could be fitted to a future experimental data, $M_{T_{4c}} = 6.9$ GeV is the mass of $T_{4c}$ tetraquark and $M_{4c}$ is the invariant mass of the $c\bar c c\bar c$-system.
In the numerical calculations we take $\Delta M = 100$ MeV.

\begin{figure}
\includegraphics[width=.6\textwidth]{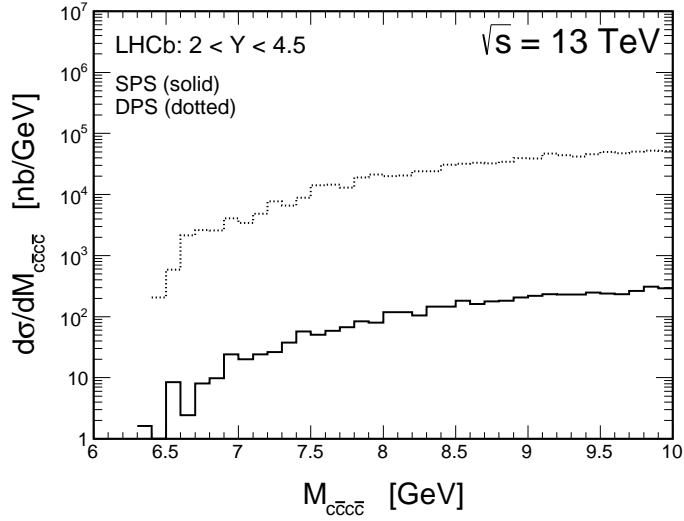}
\caption{Distribution of invariant mass of four quark-antiquark system.
Here $\sqrt{s} =$ 13 TeV and average rapidity of quarks and antiquarks
in the interval (2,4.5). The solid line is for SPS and 
the dashed line for DPS contributions.}
\label{fig:dsig_dM_4c}
\end{figure}

In Fig.\ref{fig:dsig_dM_4c} we show the invariant mass distribution
calculated for SPS (solid line) and DPS (dashed line) contribution.
In this calculation we take limitation on rapidity of the $c \bar c c \bar c$
system  2 $< Y <$ 4.5 relevant for the LHCb apparatus.
Clearly the cross section for DPS is much larger than the cross section
for SPS in the vicinity of the tetraquark position. This does not
mean that the tetraquark is produced mainly in the DPS mechanism.
The underlying production mechanism is complicated as it involves
many-body correlations and four-body wave function.  
Furthermore the production mechanism for DPS is (must be) different
than for SPS. In particular, the quarks and antiquarks produced
in the DPS mechanism may be less space-time correlated than
those from the SPS mechanism.

\begin{figure}
\includegraphics[width=.6\textwidth]{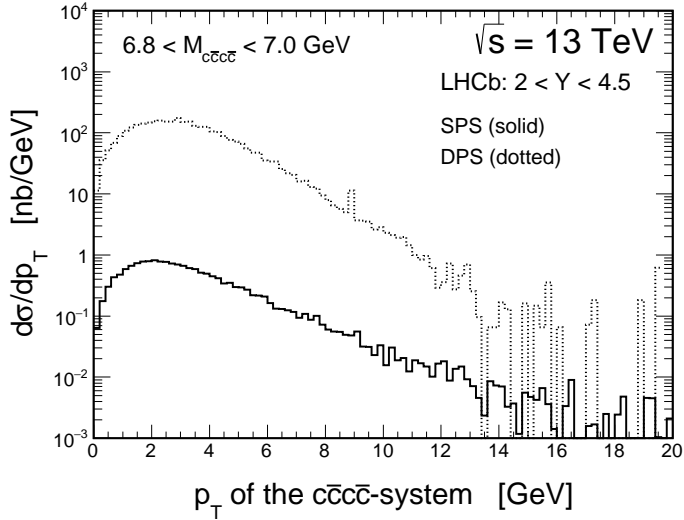}
\caption{Distribution of $p_{t,4c}$ of four quark-antiquark system
within invariant mass window $(M_R - \rm{0.1 GeV}, M_R + \rm{0.1 GeV})$.
Here $\sqrt{s} =$ 13 TeV and average rapidity of quarks and antiquarks
in the interval (2,4.5). The solid line is for SPS and the
dashed line for DPS contributions.
}
\label{fig:dsig_dpt_4c}
\end{figure}

Now we wish to visualize the $p_{t,4c}$ distribution in a very
narrow window of $M_{4c}$ in the sourrounding of the tetraquark mass.
Such a distribution is shown in Fig.\ref{fig:dsig_dpt_4c}.
This calculation requires large statistics of the experimental sample. 
Of course in general
$p_{t,4c}$ it is not $p_{t,T_{4c}}$ but must be closely related.
It would be so in a bit naive coalescence or color evaporation model.

In the present calculations we have summed over all possible color
states of outgoing $c$ quarks and $\bar c$ antiquarks.
It would be of interest to make the color study in a future.
Then, having selected the wave function of the teraquark (requires
selecting a model), one could select final states more relevant 
for the tetraquark production. We leave such a study for a future work.

\subsection{$p p \to J/\psi J/\psi$ background}

\begin{figure}
\includegraphics[width=5cm]{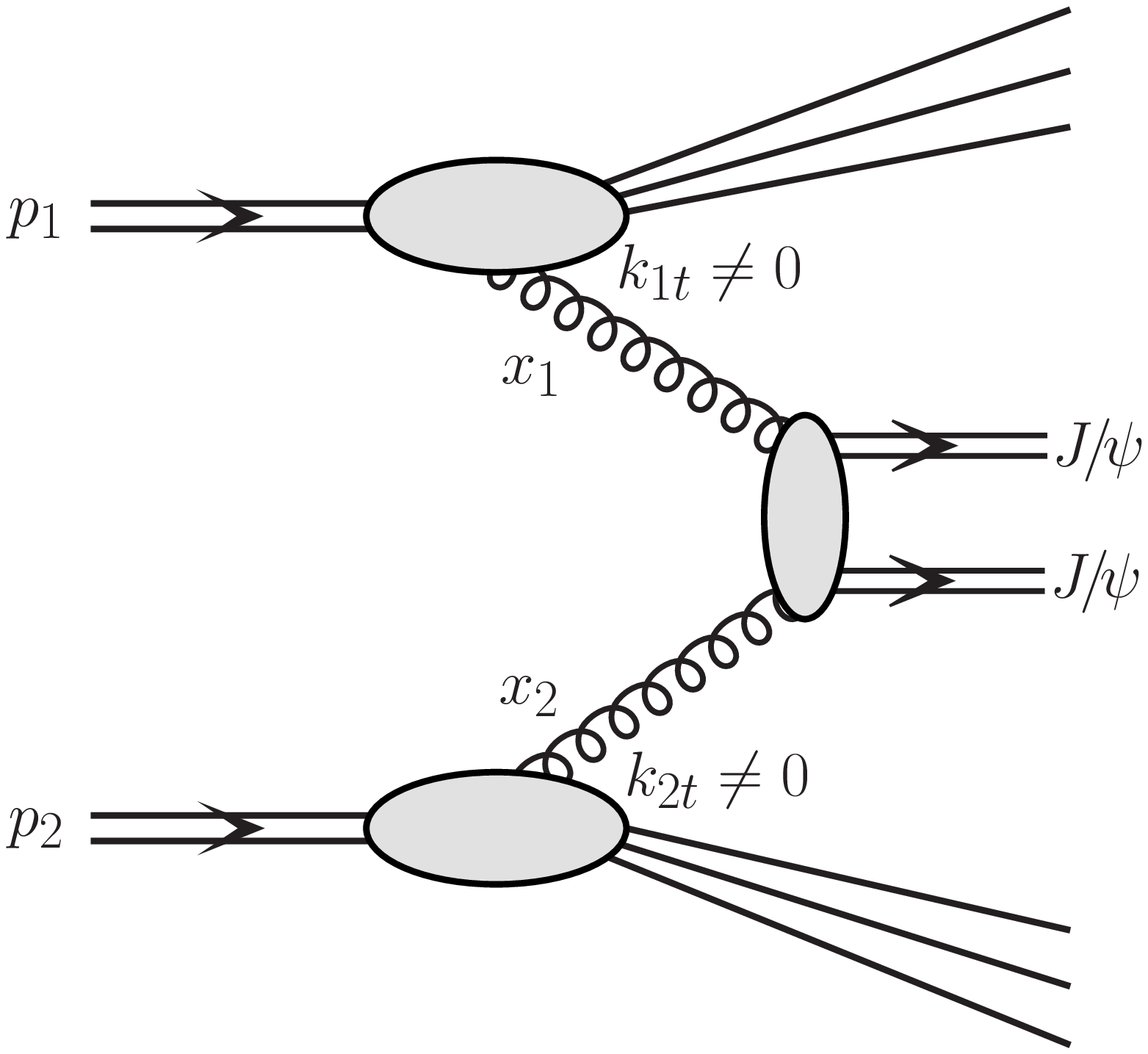}
\includegraphics[width=5cm]{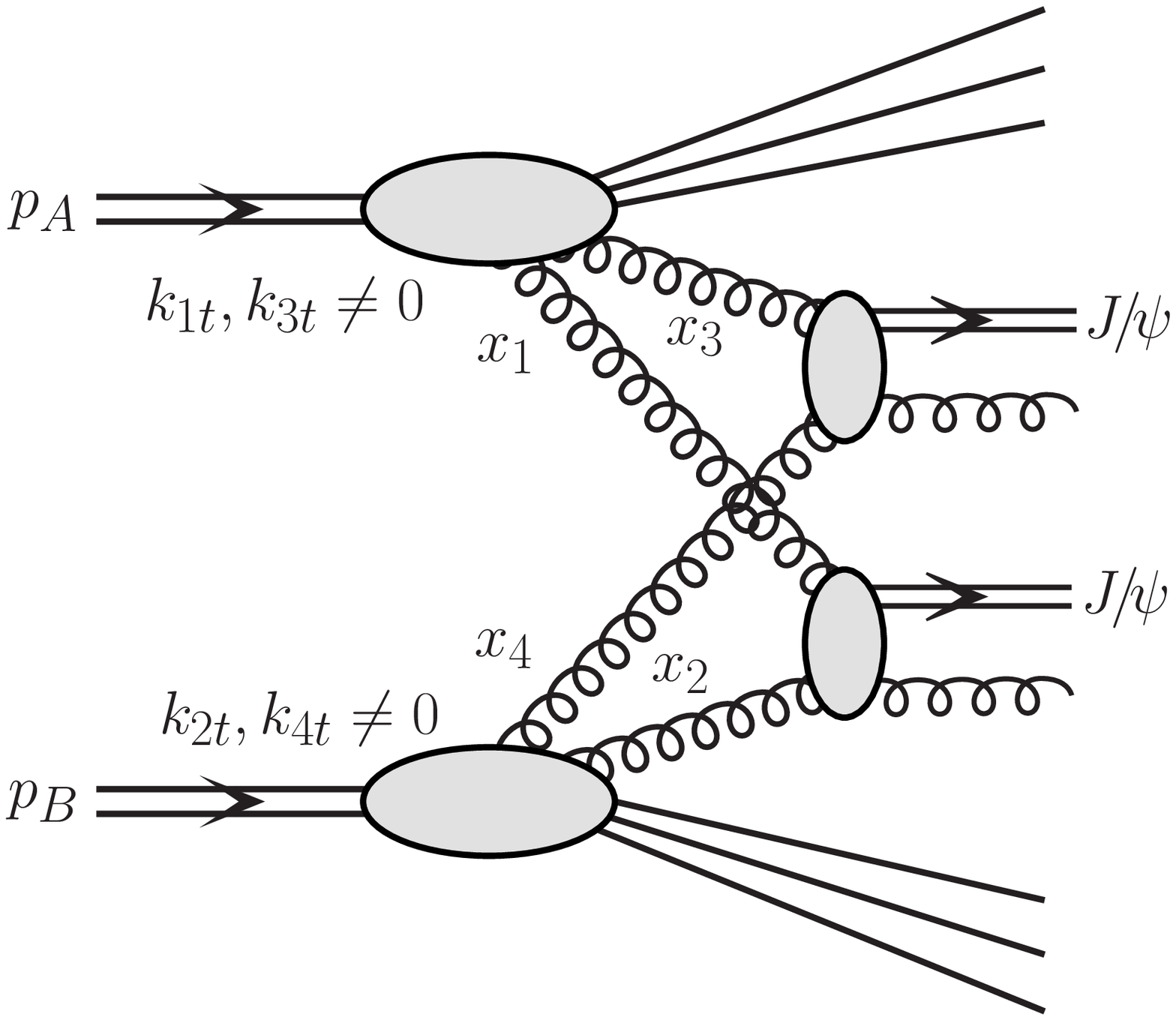}
\caption{Two dominant reaction mechanisms of production of 
$J/\psi J/\psi$ nonresonant continuum. The left diagram represent
the SPS mechanism (box type) and the right diagram the DPS mechanism.}
\label{fig:diagrams_jpsijpsi}
\end{figure}

It is of interest to calculate also background to the $J/\psi J/\psi$
final state used in the LHCb experiment. There are two dominant
mechanisms shown in Fig.\ref{fig:diagrams_jpsijpsi}.
The normalized cross section within the LHCb acceptance was
measured \cite{LHCb_jpsijpsi_13TeV}.

The single parton scattering was discussed e.g. in \cite{BLLN2011,BSZSS2013}
and the double parton scattering in \cite{KKS2011,BSZSS2013,BK2017}.
Our calculation of the SPS component is performed in $k_T$-factorization
approach analogously to Eq.(\ref{cs_formula}). We use the matrix
elements of Ref. \cite{Baranov:2011zz} for the $g g \to J/\psi J/\psi$ process.
There is also a  feeddown contribution from $g g \to J/\psi \psi'$, 
and $gg \to \psi' \psi'$ processes.
These processes are described by the same box diagrams. 
The integrated parton level cross section for $J/\psi J/\psi$
behaves as $\sim |R_{J/\psi}(0)|^4/M_{J/\psi}^8$. 
Using, $|R_{\psi'}(0)|^2/|R_{J/\psi}(0)|^2 \sim 0.4 \div 0.5$ and ${\rm Br}(\psi' \to J/\psi X) = 0.57$ as well as the average of $J/\psi$ and $\psi'$ masses for the mixed channel, one estimates an enhancement factor $\sim 1.3$ relative to the result for the $g g \to J/\psi J/\psi$ box only \cite{BLLN2011}.

Here we have considered only the dominant mechanisms. There are some
other mechansims like gluon exchange \cite{BSZSS2013}
or $\chi_c(J_1) \chi_c(J_2)$ contributions \cite{CSS2018} which are
important for the ATLAS or CMS kinematics (large $M_{J/\psi J/\psi}$,
large $\Delta y$) but negligible for the LHCb kinematics relevant for production of the tetraquark.

As far as DPS is concerned we parametrize \footnote{We do not claim here
  that this is the underlying reaction mechanism but rather an economic
  parametrization of the experimental data. Therefore our DPS
  contribution to the $J/\psi J/\psi$ channel contains effectively
all mechanisms.} 
the single $J/\psi$ production
in terms of a simple color evaporation model based on
$k_T$-factorization approach \cite{MSC2019}.
%
%
This approach is simple enough and can be nicely adjusted to the experimental
data \cite{MSC2019}. The $\sigma_{eff}$ is relatively well known
and is about 15 mb \cite{DPS_review}.
We estimate the precision of the DPS calculation at 30 \% level.

\begin{figure}
\includegraphics[width=.6\textwidth]{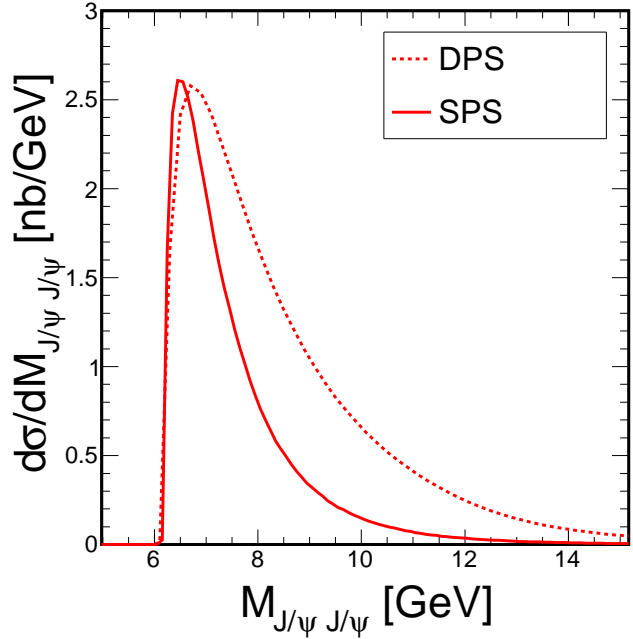}
\caption{Distribution in invariant mass of the $J/\psi J/\psi$ system
for SPS (solid line) and DPS (dashed line).
In this calculation $\sqrt{s}$ = 13 TeV and we assumed that both
$J/\psi$ mesons have rapidity in the (2,4.5) interval.}
\label{fig:dsig_dMVV}
\end{figure}

In Fig.\ref{fig:dsig_dMVV} we show distribution in $M_{J/\psi J/\psi}$
for the two mechanisms shown in Fig.\ref{fig:diagrams_jpsijpsi}.
We see that in the vicinity of the tetraquark mass the SPS contribution
is similar as the DPS one so both of them must be 
included in the evaluation of the background.

\begin{figure}
\includegraphics[width=.6\textwidth]{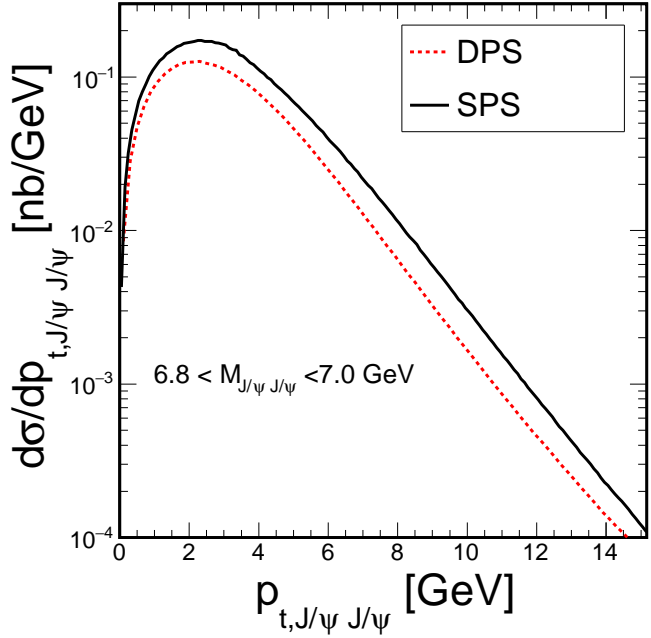}
\caption{Distribution in transverse momentum of the $J/\psi$ pairs
within the invariant mass window $(M_R - \rm{0.1 GeV}, M_R + \rm{0.1 GeV})$
for SPS (solid line) and DPS (dashed line) contributions.
Here $\sqrt{s}$ = 13 TeV.
The red lines represent the signal from the naive coalescence
approach from previous subsection multiplied by different prefactor for 
the SPS (solid line) and DPS (dashed line) $c \bar c c \bar c$ contributions.
}
\label{fig:dsig_dptsum_jpsijpsi}
\end{figure}

As for $c \bar c c \bar c$ production in the previous subsection we wish
to show distribution in $p_{t,J/\psi J/\psi}$ for the narrow window 
of invariant mass arround the tetraquark mass.
Such a distribution is shown in Fig.\ref{fig:dsig_dptsum_jpsijpsi}.
The distributions for the background here can be compared to 
the distribution of the signal from Fig.\ref{fig:dsig_dpt_4c} after
multiplying the latter by a factor 10$^{-4}$ for the DPS and 10$^{-2}$
for the SPS contributions. The $p_t$ dependence of the background
and the so-obtained signal have similar magnitude and the shape.

\subsection{$g^* g^* \to T_{4c}(6900)$ resonance production, 
examples of the spin-parity assignment}

\begin{figure}
\includegraphics[width=6cm]{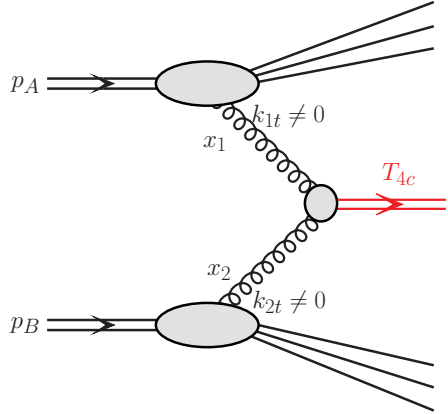}
\caption{The mechanism of gluon-gluon fusion leading to the production
of the $T_{4c}(6900)$ tetraquark.
}
\label{fig:pp_T4c}
\end{figure}

Finally we consider the calculation of the SPS-type signal
as a fusion of two (off-shell) gluons for two different spin-parity
assignments: $0^+$ and $0^-$ of the tetraquark. The corresponding
diagram is shown in Fig.\ref{fig:pp_T4c}.

In the following we use the formalism worked out recently for 
the inclusive production of pseudoscalar \cite{babiarz_pseudoscalar} 
and scalar \cite{babiarz_scalar} quarkonia. 
The off-shell gluon fusion cross sections will be proportional to a form-factor, which depends on the virtualities of gluons, $Q_i^2 = - k_i^2$:
\begin{eqnarray}
d \sigma_{g^* g^* \to 0^-} &\propto& {1 \over k_{1t}^2 k_{2t}^2} \, (\vec k_{1t} \times \vec k_{2t})^2 \, F^2(Q_1^2,Q_2^2) \nonumber \\
d \sigma_{g^* g^* \to 0^+} &\propto&  {1 \over k_{1t}^2 k_{2t}^2}  \Big( (\vec k_{1t} \cdot \vec k_{2t}) (M^2 + Q_1^2 + Q_2^2) + 2 Q_1^2 Q_2^2 \Big)^2 \,  {F^2(Q_1^2,Q_2^2) \over 4X^2} \, ,
\end{eqnarray} 
with $X = (M^4 + 2(Q_1^2+Q_2^2)M^2+(Q_1^2-Q_2^2)^2)/4$.
Note, that for the $0^+$ assignment we use only the TT coupling, as in analogy with \cite{babiarz_scalar} 
we expect the LL contribution to be smaller.
In our calculation for the tetraquark production we also use the KMR
UGDFs. 

The $g_{g g T_{4c}}$ coupling constants are in both cases roughly adjusted
to get the signal-to-background ratio of the order of 1.
In our calculation here we use the nonfactorizable monopole form factor:
\begin{equation}
F(Q_1^2,Q_2^2) = \frac{\Lambda^2}{\Lambda^2 + Q_1^2 + Q_2^2} \; ,
\end{equation}
where $Q_1^2$ and $Q_2^2$ are gluon virtualities
and vary corresponding form factor parameter $\Lambda$.
For the fully charm tetraquark one may expect naively 
$\Lambda \sim m_{T_{4c}}$ or $\Lambda \sim 4 m_c$. We will use also 
a smaller value having in mind uncertainty related to the tetraquark 
wave function.

\begin{figure}
\includegraphics[width=.49\textwidth]{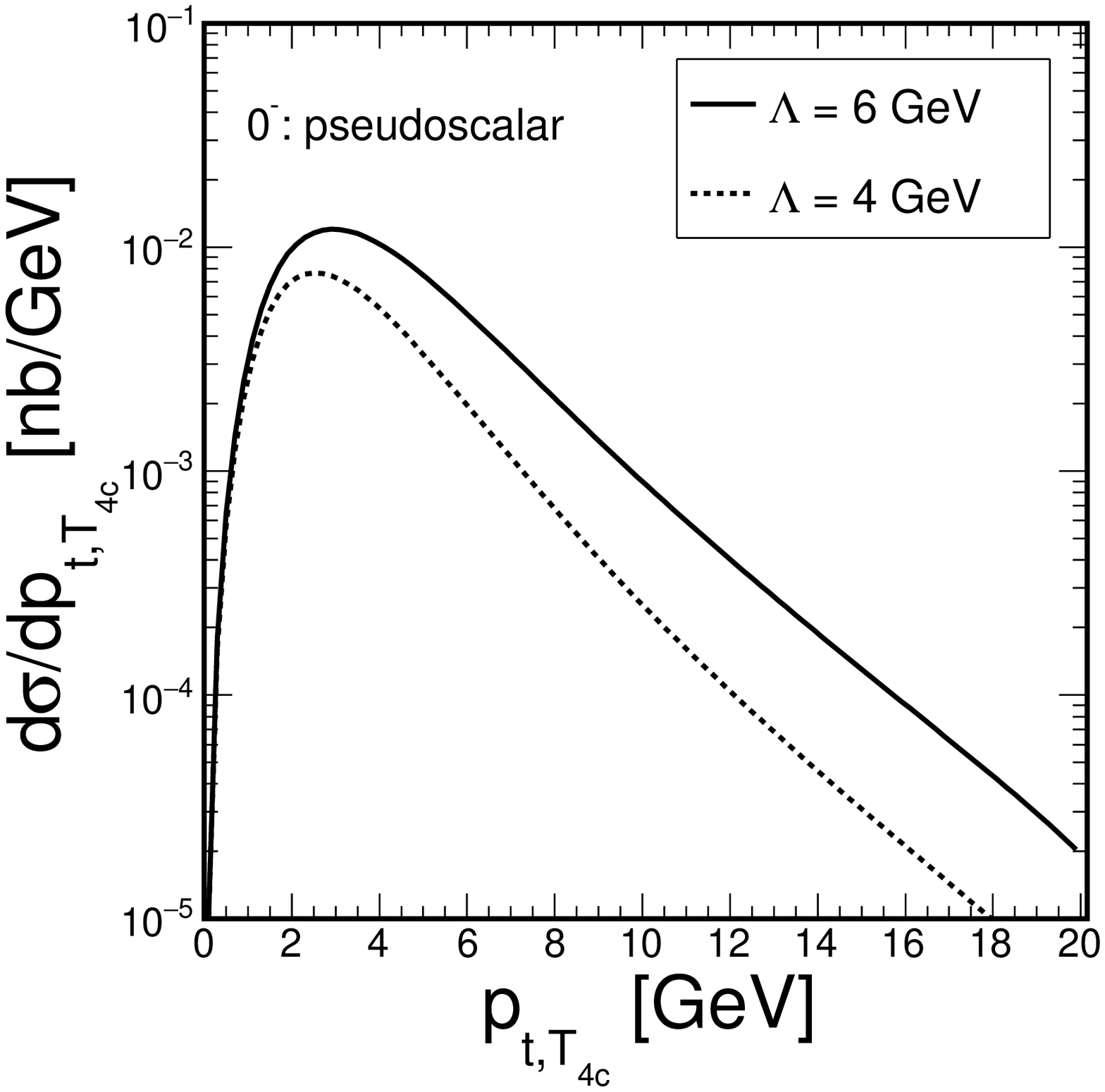}
\includegraphics[width=.49\textwidth]{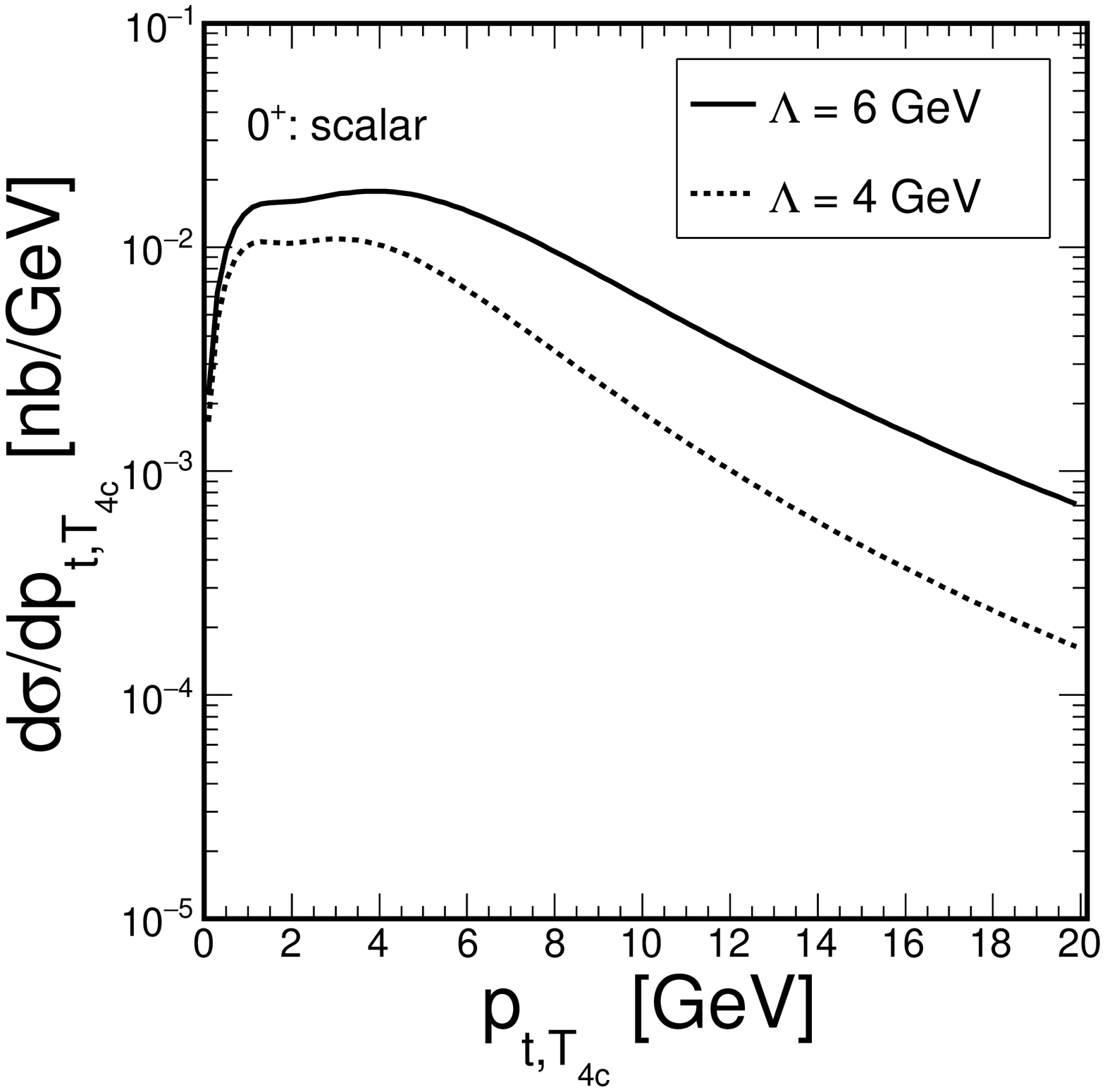}
\caption{Transverse momentum distribution of the $T_{4c}(6900)$
tetraquark for the $0^-$ (left panel) and $0^+$ (right panel)
assignments. Here $\sqrt{s}$ = 13 TeV. We show results for the
KMR UGDF and $\Lambda$ = 6 GeV (solid line) and 
$\Lambda$ = 4 GeV (dashed line).
}
\label{dsig_dpt_resonance}
\end{figure}

Since the ratio of signal-to-background improves with transverse
momentum of the tetraquark \cite{LHCb_T4c} and knowing
relatively well the behaviour of the SPS and DPS background 
(see Fig.\ref{fig:dsig_dptsum_jpsijpsi}) we can conclude that 
the $0^-$ assignment is disfavoured by the LHCb experimental results. 

Here we have considered only $0^+$ and $0^-$ ($C=+1$) spin-parity assignments.
Other assigments ($1^+$. $2^+$, etc.) should (will be) be considered 
in a future.
In fact the peak observed by the LHCb does not need to be a single 
spin but a mixture of different spins \cite{LLZZ2020,LCD2020}.
This is the main argument that we do not consider interference
of the resonance and continuum at the present stage.

\section{Conclusions}

In the present letter we have considered several aspects related to
the production of $T_{4c}(6900)$ tetraquark (called signal) observed
recently by the LHCb collaboration in the $J/\psi J/\psi$ channel and 
the $J/\psi J/\psi$ background. 
Both for the signal and the background the SPS and DPS mechanisms 
have been considered.

The background distributions can in our opinion be reliably calculated.
It is not the case for the signal. In the naive coalescence model
we have to adjust a normalization factor $C$ responsible for
the formation probability $P_{T_{4c}}$ and decay branching fraction 
$Br(T_{4c}(6900) \to J/\psi J/\psi)$ \footnote{For the branching ratio
  calculations see e.g.\cite{CCLZ2020}.}. In the moment the formation
probability cannot be calculated from first principles.
In our opinion the branching fraction is a simpler issue but also
goes beyond the scope of the present letter where we try to explore
the general situation.
Thus in the moment the product of the two unknowns can be roughly adjusted 
to the current signal-to-background ratio.
We get $C$ = 10$^{-4}$ for the DPS and C = 10$^{-2}$ for the SPS
production of $c \bar c c  \bar c$. We have not considered the mixed
scenario in which both SPS and DPS mechanisms contribute.

We have considered also more explicitly the SPS mechanism
of the resonance production via gluon-gluon fusion in 
the $k_T$-factorization approach with modern UGDFs. Also in this 
case the normalization, related to the underlying formation process
and/or wave function of the tetraquark
and the decay branching fraction $T_{4c} \to J/\psi J/\psi$ must 
be adjusted to the experimental signal-to-background ratio.
In this study we have considered two examples of the $0^+$ and $0^-$
assignment. The current data seem to exclude the $0^-$ assignment
as the final result contradicts qualitatively to the transverse 
momentum dependence of the signal-to-background ratio as observed 
by the LHCb collaboration.

In this letter we have only set the general approach leading
to a better understanding the tetraquark production. We expect that
in the future more states will be observed by the LHCb collaboration
and disentagling spins and parities will be easier.
In addition, one could study the angular correlation of the $J/\psi$ 
mesons in the tetraquark rest frame. This requires a better 
statistics available in run 3.
Then a model independent analysis \cite{MAN2020} will be possible.

In a future one could try to search for the $T_{4c}(6900)$ tetraquark
production also in the $p p \to p p J/\psi J/\psi$ exclusive reaction
or in $A A \to A A J/\psi J/\psi$ ultraperipheral collisions.
The corresponding studies will be done elsewhere.

\section*{Acknowledgments}

This study was partially supported by the Polish National Science Center
grant UMO-2018/31/B/ST2/03537 and by the Center for Innovation and
Transfer of Natural Sciences and Engineering Knowledge in Rzesz\'ow, Poland.
We are indebted to Yanxi Zhang for a discussion of some details of 
the LHCb data and Marek Karliner for a discussion of some aspects of
spectroscopy of fully charm tetraquarks.



\end{document}